\documentclass[twocolumn,showpacs,preprintnumbers,amsmath,amssymb]{revtex4}

\usepackage{graphicx}
\usepackage{dcolumn}
\usepackage{bm}
\usepackage{xcolor}

\begin{document}

\preprint{APS/123-QED}

\title{Realization of a triangular spin necklace in a verdazyl-based Ni complex}

\author{Itsuki Shimamura$^{1}$, Risa Yagura$^{1}$, Takanori Kida$^{2}$, Masayuki Hagiwara$^{2}$, Koji Araki$^{3}$, Yoshiki Iwasaki$^{4,5}$, Yuko Hosokoshi$^{1}$, Kenta Kimura$^{5,6}$, and Hironori Yamaguchi$^{1,5}$}

\affiliation{
$^1$Department of Physics, Osaka Metropolitan University, Osaka 558-8585, Japan\\
$^2$Center for Advanced High Magnetic Field Science (AHMF), Graduate School of Science, Osaka University, Osaka 560-0043, Japan\\
$^3$Department of Applied Physics, National Defense Academy, Kanagawa 239-8686, Japan\\
$^4$Department of Physics, Meisei University, Tokyo 191-8506, Japan\\
$^5$Innovative Quantum Material Center (IQMC), Osaka Metropolitan University, Osaka 558-8585, Japan\\
$^6$Department of Materials Science, Osaka Metropolitan University, Osaka 599-8531, Japan
}


Second institution and/or address\\
This line break forced

\date{\today}

\begin{abstract}
We successfully synthesized a verdazyl-based complex, ($m$-Py-V)$_3$[Ni(NO$_3$)$_2$], in which Ni$^{2+}$ ions and verdazyl radicals form a one-dimensional, triangular spin necklace consisting of spin-1/2 and spin-1 units.
Molecular orbital calculations reveal strong antiferromagnetic (AF) interactions between inversion-related radical pairs that form spin-1/2 singlet dimers. 
The remaining verdazyl and  Ni$^{2+}$  spins form frustrated triangular units, creating a distinctive spin network.
Magnetic susceptibility and specific heat measurements identify a phase transition to an AF order.
The application of magnetic fields suppresses the phase transition signal, suggesting field-induced decoupling of the spin-1 moments. 
Electron spin resonance measurements are used to evaluate the easy-axis anisotropy of spin-1, which may promote the AF order.
This work provides a rare example of a geometrically frustrated quantum spin chain realized via molecular design, thereby offering a platform for exploring frustration-driven quantum phases in low-dimensional materials.
\end{abstract}

\maketitle
\section{INTRODUCTION}
Low-dimensional quantum spin systems continue to serve as fertile platforms for uncovering exotic quantum phases driven by enhanced quantum fluctuations and strong correlations. 
Among these, one-dimensional (1D) spin chains have played a central role in the exploration of unconventional ground states and collective excitations, providing deep insights into the nature of quantum critical behavior and symmetry-protected topological phases~\cite{TLL,Haldane}.
When geometric frustration is introduced into such 1D systems, the interplay between the lattice topology and exchange interactions gives rise to even richer physics, including spontaneous dimerization and magnetization plateaus. 
Among the various 1D frustrated spin chains, zigzag chains and diamond chains have been extensively investigated both theoretically and experimentally as paradigmatic systems~\cite{zig1,zig2,zig3,zig4,zig5, dia1,dia2,dia3,dia4}. 
These models capture the frustration-induced essential physics and exhibit diverse phenomena.
In contrast, many other theoretically proposed 1D frustrated spin chains remain largely unexplored in experiments due to the absence of suitable model materials~\cite{ta1,ta2,ta3,ta4,ta5,ta6,ta7,ta8,ta9,ta10}. 
This gap between theory and experiment has posed a major obstacle to advancing our understanding of frustration-induced quantum phenomena in 1D systems. 
Consequently, the development of new quantum spin models through deliberate structural design has been a key strategy.

In addition to theoretical advancements, recent progress in molecular and crystal design has enabled the construction of spin systems with precisely controlled interaction geometries and dimensionalities. 
In particular, organic radical-based materials provide a highly tunable platform for engineering quantum spin models, where magnetic interactions can be tailored through the rational design of molecular structures and crystal packing. 
Using this approach, we have successfully realized various geometrically frustrated spin systems. 
These have included two-dimensional frustrated square lattices~\cite{TCNQ_square,PF6,SbF6,Zn_gap,MnCl4}, zigzag–square networks~\cite{zigzag_square}, and even three-dimensional quantum pentagonal lattices~\cite{a26Cl2V}, each embodying distinct frustration mechanisms and quantum behaviors.
These achievements demonstrate the potential of molecular-based materials as programmable hosts for exotic spin models, thus bridging the gap between theoretical constructs and real quantum materials.

Through molecular-based material design, we recently succeeded in realizing a spin-1/2 Kondo necklace model~\cite{Kondo}, which can be viewed as a simplified version of the Kondo lattice, focusing solely on the spin degree of freedom~\cite{Doniach}. 
In this system, Kondo-type exchange interactions between the spin chain and decorated spins stabilize the quantum singlet state. 
When a magnetic field is applied, the system exhibits a characteristic decoupling behavior, in which the decorated spins become polarized and disentangle from the chain, resulting in a field-induced phase transition. 
Furthermore, numerical studies have suggested that incorporating anisotropy, frustration, or higher dimensionality into such models could lead to even more exotic quantum phases~\cite{riron1,riron2,riron4}.

Motivated by this perspective, we extended the Kondo necklace framework by introducing geometric frustration into the spin network. 
The triangular spin necklace introduced in this study features a localized spin-1/2 coupled to two additional spins—a spin-1 and another spin-1/2—forming a triangular unit. 
The resulting triangular geometry introduces local frustration into the chain. 
This design realizes a class of quantum spin systems that combines spin imbalance, frustration, and asymmetric interactions within a controllable molecular framework.

In this study, we synthesized ($m$-Py-V)$_3$[Ni(NO$_3$)$_2$] ($m$-Py-V = 3-(3-pyridinyl)-1,5-diphenylverdazyl), a verdazyl-based complex. 
Molecular orbital calculations revealed strong antiferromagnetic (AF) interactions between inversion-related verdazyl radicals, leading to singlet dimer formation and leaving residual spins to form a frustrated triangular spin necklace with Ni$^{2+}$ ions.
Magnetic susceptibility and specific heat measurements identified a phase transition to an AF order.
The application of a magnetic field suppressed the phase transition signal, suggesting field-induced decoupling of the spin-1 moments. 
The magnetization curve confirmed the full polarization of the spins forming the triangular spin necklace above approximately 4 T. 
Electron spin resonance (ESR) measurements were performed to evaluate the easy-axis anisotropy associated with spin-1.
This anisotropy is considered to stabilize the ordered state of spin-1, thereby stabilizing the AF order throughout the system via exchange couplings with spin-1/2.

\section{EXPERIMENTAL}
We synthesized $m$-Py-V via the conventional procedure for producing the verdazyl radical~\cite{verd}.
A solution of Ni(NO$_3$)$_2$$\cdot$6H$_2$O (58 mg, 0.2 mmol) in 2 ml ethanol was slowly added to a solution of $m$-Py-V (189 mg, 0.6 mmol) in 6 ml of CH$_2$Cl$_2$ and stirred for 30 min. 
A dark-green crystalline solid of ($m$-Py-V)$_3$[Ni(NO$_3$)$_2$]  was separated by filtration. 
Single crystals were obtained via recrystallization from CH$_2$Cl$_2$ at 10 $^\circ$C.

The X-ray intensity data were collected using a Rigaku XtaLAB Synergy-S instrument.
Anisotropic and isotropic thermal parameters were employed for non-hydrogen and hydrogen atoms, respectively, during the structure refinement. 
The hydrogen atoms were positioned at their calculated ideal positions.
Magnetization measurements were conducted using a commercial SQUID magnetometer (MPMS, Quantum Design).
The experimental results were corrected by considering the diamagnetic contributions calculated using Pascal's method.
High-field magnetization in pulsed magnetic fields was measured using a non-destructive pulse magnet.
Specific heat measurements were performed using a commercial calorimeter (PPMS, Quantum Design) employing a thermal relaxation method.
The ESR measurements were performed utilizing a vector network analyzer (ABmm) and a superconducting magnet (Oxford Instruments).
All the experiments utilized small, randomly oriented single crystals.

Molecular orbital (MO) calculations were performed using the UB3LYP method.
The basis sets are 6-31G (intermolecule) and 6-31G($d$, $p$) (intramolecule).
All calculations were performed using the GAUSSIAN09 software package.
The convergence criterion was set at 10$^{-8}$ hartrees.
We employed a conventional evaluation scheme to estimate the intermolecular exchange interactions in the molecular pairs~\cite{MOcal}. 


\begin{figure*}[t]
\begin{center}
\includegraphics[width=40pc]{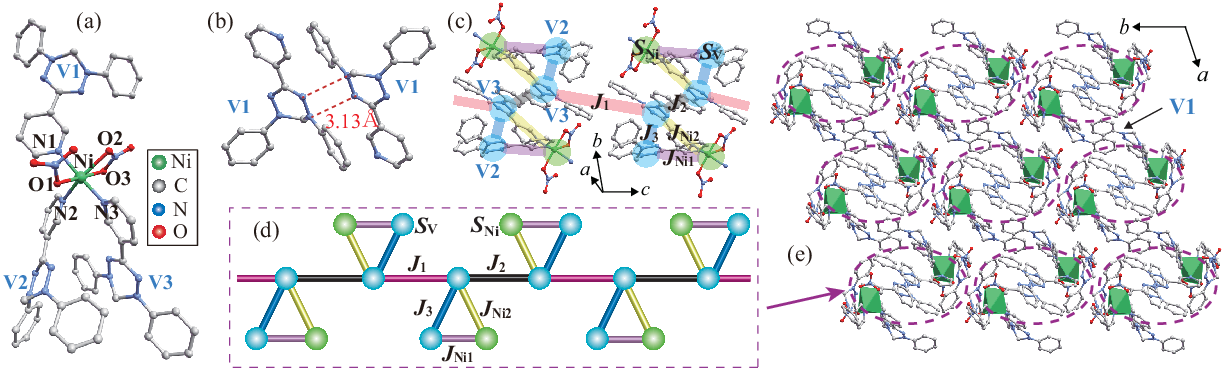}
\caption{(color online) (a) Molecular structure of ($m$-Py-V)$_3$[Ni(NO$_3$)$_2$]. 
The hydrogen atoms have been omitted for clarity. 
(b) Molecular pair associated with the exchange interaction $J_0$.
(c) Crystal structure forming a triangular spin necklace along the $c$ axis.
The blue and green nodes represent the spin-1/2 on the radical and the spin-1 on the Ni ion, $S_{\rm{V}}$ and $S_{\rm{Ni}}$, respectively. 
The thick lines represent the exchange interactions. 
(d) Corresponding triangular spin necklace.
(e) Crystal structure in the $ab$ plane.
The broken line encloses the molecules comprising each triangular spin necklace along the $c$ axis. 
}
\end{center}
\end{figure*}

\section{RESULTS AND DISCUSSION}
\subsection{Crystal structure and spin model}
The crystallographic parameters of ($m$-Py-V)$_3$[Ni(NO$_3$)$_2$] are listed in Table I.
Figure 1(a) shows its molecular structure, where the verdazyl radical, $m$-Py-V, and Ni$^{2+}$ have spin values of 1/2 and 1, respectively.
The Ni$^{2+}$ ion is coordinated with N atoms from three $m$-Py-V ligands and O atoms from two nitrate ligands, resulting in an octahedral coordination environment. 
Among the two nitrate ligands, one acts as a monodentate ligand coordinating through a single O atom, whereas the other functions as a bidentate ligand coordinating through two O atoms.
Table II lists the bond lengths and angles relevant to the Ni atom. 
Regarding the spin density distribution in the radicals, MO calculations showed that $\sim $61 ${\%}$ of the total spin density was localized on the central ring consisting of four N atoms, whereas each phenyl ring directly attached to the central N atom contributed $\sim $16 ${\%}$ of the spin density.
The dominant exchange interactions were identified through the MO calculations.
We evaluated a strong AF interaction between the radicals labeled V1, which is related by the inversion symmetry and has an N-N short contact, as shown in Fig. 1(b).
Its value was evaluated as $J_0$/$k_{\rm{B}}$= 572 K, which is defined in the Heisenberg spin Hamiltonian given by $\mathcal {H} = J_{n}{\sum^{}_{<i,j>}}\textbf{{\textit S}}_{i}{\cdot}\textbf{{\textit S}}_{j}$. 
Because this strong AF interaction is expected to form a nonmagnetic singlet state at low temperatures, an effective spin model is developed from the exchange interactions related to the radicals labeled V2 and V3 in Fig. 1(a).
We found three types of dominant interactions $J_1$/$k_{\rm{B}}$ = 4.9 K,  $J_2$/$k_{\rm{B}}$  = $-$4.8 K, and $J_3$/$k_{\rm{B}}$ = 2.0 K between radical spins, as shown in Fig. 1(c).
Two types of ferromagnetic (F) interactions between Ni and radical spins were evaluated: $J_{\rm{Ni1}}/k_{\rm{B}}$ = $-$11.6 K and $J_{\rm{Ni2}}/k_{\rm{B}}$ = $-$11.2 K.  
Given that MO calculations tend to overestimate the intramolecular interactions involving transition metals, the actual values are typically approximately half of the calculated values~\cite{morotaMn,tominagaNi}. 
This situation results in an energy scale comparable to that of $J_1$–$J_3$ interactions.
Consequently, assuming all the expected interactions, a triangular spin necklace comprising $S_{\rm{V}}$=1/2 and $S_{\rm{Ni}}$=1 is formed along the $c$ axis, as shown in Figs. 1(c) and 1(d).
The 1D alternating chain is formed by $J_{\rm{1}}$ and $J_{\rm{2}}$, and an additional spin-1/2 and spin-1 are decorated via $J_{\rm{3}}$ and $J_{\rm{Ni2}}$, respectively. 
Furthermore, $J_{\rm{3}}$, $J_{\rm{Ni1}}$, and $J_{\rm{Ni2}}$ form a triangular unit with frustration caused by one AF and two F interactions.
The V1 radical pairs that form a nonmagnetic singlet via $J_0$ exist between the 1D structures, as shown in Fig. 1(e). 
Hence, the exchange interactions caused by the overlapping of the MOs between the spin necklaces are expected to be weak.

\begin{table}
\caption{Crystallographic data of ($m$-Py-V)$_3$[Ni(NO$_3$)$_2$].}
\label{t1}
\begin{center}
\begin{tabular}{lc}
\hline
\hline 
Formula & C$_{57}$H$_{48}$N$_{17}$NiO$_{6}$\\
Crystal system & Monoclinic \\
Space group & $P$$\Bar{1}$ \\
Temperature (K) & 100 \\
$a$ $(\rm{\AA})$ & 13.9237(3) \\
$b$ $(\rm{\AA})$ & 14.9493(4) \\
$c$ $(\rm{\AA})$ & 15.0912(3) \\
$\alpha$ (degrees) &  93.2197(19) \\
$\beta$ (degrees) &  115.052(2) \\
$\gamma$ (degrees) &  106.281(2) \\
$V$ ($\rm{\AA}^3$) & 2675.79(11) \\
$Z$ & 2 \\
$D_{\rm{calc}}$ (g cm$^{-3}$) & 1.397\\
Total reflections & 6682 \\
Reflection used & 5781 \\
Parameters refined & 730 \\
$R$ [$I>2\sigma(I)$] & 0.0351 \\
$R_w$ [$I>2\sigma(I)$] & 0.0909 \\
Goodness of fit & 1.028 \\
CCDC & 2457181\\
\hline
\hline
\end{tabular}
\end{center}
\end{table}

\begin{table}
\caption{Bond lengths ($\rm{\AA}$) and angles ($^{\circ}$) related to the Ni atom in ($m$-Py-V)$_3$[Ni(NO$_3$)$_2$].}
\label{t1}
\begin{center}
\begin{tabular}{cc@{\hspace{1.5cm}}cc}
\hline
\hline
Ni--N1 & 2.09 & O1--Ni--O2 & 115.7 \\
Ni--N2 & 2.07 & O2--Ni--O3 & 59.9 \\
Ni--N3 & 2.08 & O3--Ni--N2 & 95.9 \\
Ni--O1 & 2.05 & N2--Ni--O1 & 88.4 \\
Ni--O2 & 2.21 & O1--Ni--N1 & 86.6 \\
Ni--O3 & 2.10 & N1--Ni--O2 & 89.3 \\
 &  & O2--Ni--N3 & 86.8 \\
 &  & N3--Ni--O1 & 89.9 \\
 &  & N1--Ni--O2 & 89.3 \\
 &  & O2--Ni--N3 & 86.8 \\
 &  & N3--Ni--N2 & 93.6 \\
 &  & N2--Ni--N1 & 92.3 \\
\hline
\hline
\end{tabular}
\end{center}
\end{table}

\begin{figure}[t]
\begin{center}
\includegraphics[width=21pc]{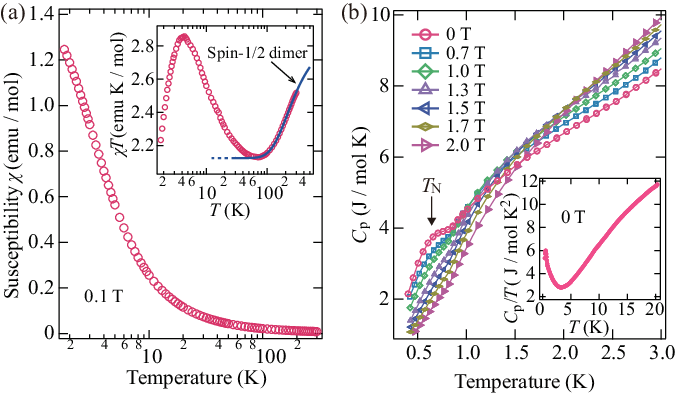}
\caption{(color online)  Temperature dependence of (a) the magnetic susceptibility ($\chi$ = $M/H$) and (b) $\chi$$T$ of ($m$-Py-V)$_3$[Ni(NO$_3$)$_2$] at 0.1 T. 
The solid line represents the result calculated for a spin-1/2 AF dimer via $J_{\rm{0}}$.
(b) Temperature dependence of the specific heat $C_{\rm{p}}$ of ($m$-Py-V)$_3$[Ni(NO$_3$)$_2$].
The inset shows $C_{\rm{p}}$/$T$ at zero field with the expanded temperature regime.
}\label{f1}
\end{center}
\end{figure}

\subsection{Magnetic susceptibility}
Figure 2(a) shows the temperature dependence of the magnetic susceptibility ($\chi$) and $\chi T$ of the complex at 0.1~T.
The $\chi T$ exhibits a steep decrease with decreasing temperature down to $\sim$70 K, indicating the formation of a singlet dimer of spin-1/2 coupled by the strong AF $J_0$, as shown in the inset of Fig. 2(a). 
The effective interactions through the dimer state for $T \ll$ $J_0$/$k_{\rm{B}}$ are expected to be negligible~\cite{Masuda,b26Cl2V,random_ladder}. 
Accordingly, the exchange interactions that form the triangular spin necklace are expected to be effective in sufficiently low temperature regions. 
Then, the increase in $\chi T$ down to $\sim$4~K demonstrates the dominant contributions of F interactions, i.e., $J_{\rm{2}}$, $J_{\rm{Ni1}}$, and $J_{\rm{Ni2}}$. 
The subsequent decrease in $\chi T$ below $\sim$4~K reflects the effects of weaker AF interactions, i.e., $J_{\rm{1}}$ and $J_{\rm{3}}$, which give rise to the peak in $\chi T$.
The entire $\chi$ down to the lowest experimental temperature exhibits paramagnetic-like behavior, reflecting the small energy scale of the interactions and the competition between AF and F interactions.

We calculated the magnetic susceptibility of the spin-1/2 AF dimer and fitted it to the experimental $\chi$$T$ data above $\sim$70 K.
The calculated value was shifted up by 2.13~emu·K/mol, which corresponds to the expected Curie constant for $S_{\rm{V}}$ and $S_{\rm{Ni}}$ forming the triangular spin necklace, assuming $g$-factors of 2.0 for $S_{\rm{V}}$ and approximately 2.2 for $S_{\rm{Ni}}$.
The experimental behavior was then explained using $J_{0}/k_{\rm{B}}$ = 456(1) K, which is consistent with the energy scale evaluated from the MO calculations, as shown in the inset of Fig. 2(a).

\subsection{Specific heat}
The temperature dependence of the specific heat $C_{\rm{p}}$ is shown in Fig. 2(b). 
The magnetic contributions are expected to be dominant in the low-temperature regions considered here. 
At zero field, we found an anomalous change at $T_{\rm{N}}$ = 0.65 K, which remains almost filed independent up to approximately 0.7 T.
This anomaly is attributed to a phase transition into an AF ordered state induced by weak but finite interchain couplings.
In low-dimensional spin systems, the development of short-range correlations significantly reduces the entropy change at the transition, so that the anomaly in the specific heat does not necessarily appear as a sharp $\lambda$-type peak.
Actually, the broad anomaly observed around 1.0–2.0~K is considered to originate from short-range correlations within the 1D chain.
Although interchain interactions are expected to be weak based on the molecular packing, the observed $T_{\rm{N}}$ value is relatively high compared to the energy scale of the dominant intrachain couplings. 
This finding implies that magnetic anisotropy also contributes to stabilization of the AF order. 
The presence of anisotropy is further supported by ESR measurements, which will be discussed later.
The peak at zero field decreases with increasing magnetic field and almost disappears. 
This behavior is different from that of the conventional AF order, in which $T_{\rm{N}}$ typically shifts to lower temperatures under applied magnetic fields. 
Instead, the observed suppression resembles the decoupling behavior reported in the Kondo necklace model, where large-moment spins coupled via Kondo couplings decouple from the chain in the presence of a magnetic field~\cite{Kondo}. 
In the present case, the field-induced disappearance of the peak is considered to arise from the decoupling of $S_{\rm{Ni}}$ in the triangular unit.
As shown in the inset of Fig. 2(b), the temperature dependence of $C_{\rm{p}}/T$ indicates that additional lower-temperature data would be required to fully evaluate the entropy change associated with the phase transition via integration of $C_{\rm{p}}/T$.

\subsection{Magnetization curve}
Figure 3 shows the magnetization curve at 1.4 K measured in pulsed magnetic fields, with its inset showing the low-field region measured in static magnetic fields at 1.8 K. 
The magnetization gradually increases up to $\sim$4.27~$\mu_{\rm{B}}$/f.u., which persists up to at least 51 T. 
Because the AF dimer coupled by $J_{0}$ has a large spin gap beyond 300 T, the magnetic behavior in the present field region is attributed to the triangular spin necklace.     
The observed result is reminiscent of paramagnetic-like behavior, which is due to the small energy scale of the exchange interactions and is consistent with the temperature dependence of magnetic susceptibility. 
The value of 4.27~$\mu_{\rm{B}}$/f.u. corresponds to the full polarization of both V2 and V3 spins $S_{\rm{V}}$ (=1/2) with an isotropic $g$ value of 2.00 and $S_{\rm{Ni}}$ (=1) with a $g$ value of 2.27 in the magnetic unit cell, i.e., 2~×(2.00×1/2)+2.27×1=4.27.
In anisotropic spin systems, unfixed powder samples tend to align along the direction of an external magnetic field~\cite{Kondo,morotaCo}. 
Such field-induced alignment allows measurements to reflect the response along the easy axis.
Considering that the observed field-polarized value of the magnetization indicates the contribution from $S_{\rm{Ni}}$ with the $g$-value for the easy axis, as evaluated from the ESR analysis, field-induced alignment was also expected in the present experiment.
If the dominant F interactions, $J_{\rm{Ni1}}$ and $J_{\rm{Ni2}}$, stabilized a fully correlated state, an effective spin-2 moment would be expected at low temperatures.  
However, the observed magnetization behavior is more accurately reproduced by the Brillouin function, assuming independent spin-1 and spin-1/2 moments, rather than a single spin-2 moment, as shown in the inset of Fig.3.  
This result supports the scenario in which the spin-1 ($S_{\rm{Ni}}$) decouples from the spin-1/2 system under magnetic fields, consistent with the interpretation based on the specific heat measurements.

\begin{figure}[t]
\begin{center}
\includegraphics[width=18pc]{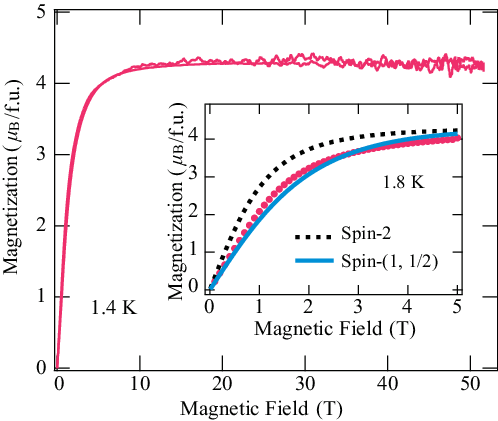}
\caption{(color online) Magnetization curves of ($m$-Py-V)$_3$[Ni(NO$_3$)$_2$] at 1.4 K under applied pulsed magnetic fields.
The inset shows the low-field region measured in static magnetic fields at 1.8 K. The broken and solid lines represent the
Brillouin function for spin-2 and spin-(1/2, 1), respectively. 
}\label{f1}
\end{center}
\end{figure}

\subsection{Electron spin resonance}
We performed ESR measurements to examine the magnetic anisotropy of $S_{\rm{Ni}}$. 
Figure 4(a) shows the frequency dependences of the resonance signals at 1.8 K. 
Because the experiments were performed using powder samples fixed with grease, the observed signals corresponded to the resonance fields for the external field parallel to the principal axes. 
The broad field range of the resonance signal necessitated the use of high-frequency, high-field conditions without a cavity. 
Under these conditions, the transmitted signal intensity is inherently weak, resulting in a relatively low signal-to-noise ratio.
We evaluated intrinsic resonance signals from $S_{\rm{Ni}}$ with $g$$\sim$ 2.2 and plotted the resonance fields in the frequency-field diagram, as shown in Fig. 4(b).
Assuming the spin-1 monomer, we consider the on-site anisotropy as $\mathcal {H} = D(S_{z})^2-\mu_{\rm{B}}\textbf{{\textit H}}\bf{\tilde  g}\textbf{{\textit S}}$, where $\mu_{B}$ is the Bohr magneton and $\bf{\tilde  g}$ denotes the $g$-tensor.
The diagonal components of the principal axes of the $g$-tensor are $g_x$, $g_y$, and $g_z$, and the other components are zero.
Figure 4(c) shows the energy levels calculated using the evaluated parameters. 
The resonance modes at a sufficiently low temperature of 1.8~K correspond to the transitions indicated by the arrows in the energy branches. 
As shown in Fig. 4(b), we obtained good agreement between the experimental and calculated results for the resonance modes using $D/k_{\rm{B}}$=$-$1.3~K, $g_x$=$g_y$=2.22 ($H\perp z$), and $g_z$=2.27 ($H//z$). 
The estimated $g$ values are consistent with the magnetization measurements, supporting the reliability of the ESR analysis. 
Moreover, the easy-axis anisotropy of $S_{\rm{Ni}}$ is expected to stabilize the ordered state associated with $S_{\rm{Ni}}$, leading to the AF order in the entire system through exchange couplings $J_{\rm{Ni1}}$ and $J_{\rm{Ni2}}$ at zero field.
Importantly, in magnetic fields, the observed ESR signals are well explained by the spin-1 monomer model.  
This agreement provides further experimental support for the field-induced decoupling scenario of $S_{\rm{Ni}}$ in the triangular spin necklace.
If $J_{\rm{Ni1}}$ and $J_{\rm{Ni2}}$ formed a strongly coupled trimer, the ground state would be an effective spin-2, and ESR signals would appear at fields corresponding to an averaged $g$ value between 2.0 and 2.2.  
However, the observed signals are well explained by a spin-1 monomer with $g \approx 2.2$, which contradicts the trimer scenario.

\begin{figure}[t]
\begin{center}
\includegraphics[width=19pc]{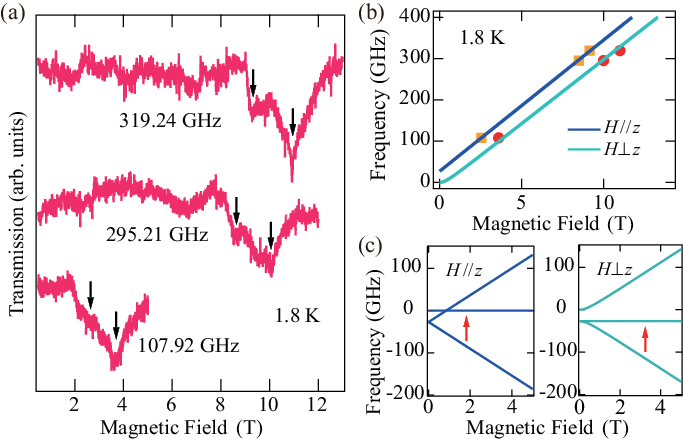}
\caption{(color online) 
(a) Frequency dependence of ESR absorption spectra of ($m$-Py-V)$_3$[Ni(NO$_3$)$_2$] at 1.8 K. 
The arrows indicate the resonance fields.
(b) Frequency-field plot of the resonance fields.
Solid lines indicate the calculated resonance modes of the spin-1 monomer with on-site anisotropy.
(c) Calculated energy branch of the the spin-1 monomer for $H$//$z$ and $H\perp z$.
Arrows indicate spin-allowed transitions from the ground state, which correspond to the resonance modes.
}\label{f3}
\end{center}
\end{figure}

\section{Summary}
In summary, we synthesized a verdazyl-based complex, ($m$-Py-V)$_3$[Ni(NO$_3$)$_2$].
The Ni$^{2+}$ ion coordinates with three $m$-Py-V ligands and two nitrate ligands, resulting in an octahedral coordination environment. 
MO calculations indicated a strong AF interaction between the radical pair related by inversion symmetry, leading to a nonmagnetic singlet state at low temperatures.
The residual radical and Ni spins formed a triangular spin necklace composed of spin-1/2 and spin-1 with frustration.
In the low-temperature region, the magnetic susceptibility demonstrated that a dominant F contribution and subsequent weaker AF contributions appeared with decreasing temperature.
The specific heat exhibited an anomalous change at $T_{\rm{N}}$ = 0.65 K, indicating AF order stabilized by weak but finite interchain couplings.  
The anomaly was suppressed by an external field and eventually disappeared, suggesting the field-induced decoupling of the spin-1 in the triangular units.
The magnetization curve revealed a gradual increase, followed by asymptotic behavior to a constant value, indicating full polarization of both spin-1/2 and spin-1 in the triangular spin necklace.  
High-frequency ESR measurements revealed an easy-axis anisotropy of spin-1.
This anisotropy is expected to stabilize the ordered state of spin-1, contributing to the stabilization of the AF order throughout the system via coupling with spin-1/2.
This study demonstrates the feasibility of a molecular-based design for constructing geometrically frustrated 1D spin systems with well-controlled interactions.  
The realization of a triangular spin necklace offers a platform for investigating frustration-driven quantum phenomena in low-dimensional materials.

\begin{acknowledgments}
This research was partly supported by KAKENHI (Grants No. 24K00575 and No. 23K25824) and the Shorai Foundation for Science and Technology.
A part of this work was performed under the interuniversity cooperative research program of the joint-research program of ISSP, the University of Tokyo.
\end{acknowledgments}

The data that support the findings of this article are openly available~\cite{data}.



\begin{thebibliography}{99}

\bibitem{TLL}
S. Tomonaga, Remarks on Bloch's method of sound waves applied to many-fermion problems, Prog. Theor. Phys. \textbf{5}, 544 (1950).

\bibitem{Haldane}
F. D. M. Haldane, Nonlinear field theory of large-spin Heisenberg antiferromagnets: semiclassically quantized solitons of the one-dimensional easy-axis Néel state, Phys. Rev. Lett. \textbf{50}, 1153 (1983).


\bibitem{zig1}
N. Shannon, T. Momoi, and P. Sindzingre, Nematic order in square lattice frustrated ferromagnets, Phys. Rev. Lett. \textbf{96}, 027213 (2006).

\bibitem{zig2}
T. Vekua, A. Honecker, H.-J. Mikeska, and F. Heidrich-Meisner, Correlation functions and excitation spectrum of the frustrated ferromagnetic spin-1/2 chain in an external magnetic field, Phys. Rev. B \textbf{76}, 174420 (2007).

\bibitem{zig3}
T. Hikihara, L. Kecke, T. Momoi, and A. Furusaki, Vector chiral and multipolar orders in the spin-1/2 frustrated ferromagnetic chain in magnetic field, Phys. Rev. B \textbf{78}, 144404 (2008).

\bibitem{zig4}
L. E. Svistov, T. Fujita, H. Yamaguchi, S. Kimura, K. Omura, A. Prokofiev, A. I. Smirnov, Z. Honda, and M. Hagiwara, New high magnetic field phase of the frustrated $S$ = 1/2 chain compound LiCuVO$_4$, JETP Lett. \textbf{93}, 21 (2011).

\bibitem{zig5}
M. Pregelj, A. Zorko, O. Zaharko, H. Nojiri, H. Berger, L. Chapon, and D. Ar$\check{\rm{c}}$on, Spin-stripe phase in a frustrated zigzag spin-1/2 chain, Nat. Commun. \textbf{6}, 7255 (2015).

\bibitem{dia1}
K. Okamoto, T. Tonegawa, and M. Kaburagi, Magnetic properties of the $S$=1/2 distorted diamond chain at $T$ = 0, J. Phys.: Condens. Matter \textbf{15}, 5979 (2003).

\bibitem{dia2}
H. Kikuchi, Y. Fujii, M. Chiba, S. Mitsudo, T. Idehara, T. Tonegawa, K. Okamoto, T. Sakai, T. Kuwai, and H. Ohta, Experimental observation of the 1/3 magnetization plateau in the diamond-chain compound Cu$_3$(CO$_3$)$_2$(OH)$_2$, Phys. Rev. Lett. \textbf{94}, 227201 (2005).

\bibitem{dia3}
B. Gu and G. Su, Magnetism and thermodynamics of spin-1/2 Heisenberg diamond chains in a magnetic field, Phys. Rev. B \textbf{75}, 174437 (2007).

\bibitem{dia4}
J. Stre$\check{\rm{c}}$ka, T. Verkholyak, J. Richter, K. Karlov$\acute{\rm{a}}$, O. Derzhko, and J. Schnack, Frustrated magnetism of spin-1/2 Heisenberg diamond and octahedral chains as a statistical mechanical monomer-dimer problem, Phys. Rev. B \textbf{105}, 064420 (2022).


\bibitem{ta1}
P. Azaria, C. Hooley, P. Lecheminant, C. Lhuillier, and A. M. Tsvelik, Kagom$\acute{\rm{e}}$ lattice antiferromagnet stripped to its basics, Phys. Rev. Lett. \textbf{81}, 1694 (1998).

\bibitem{ta2}
J. Schulenburg, A. Honecker, J. Schnack, J. Richter, and H.-J. Schmidt, Macroscopic magnetization jumps due to independent magnons in frustrated quantum spin lattices, Phys. Rev. Lett. \textbf{88}, 167207 (2002).

\bibitem{ta3}
T. Shimokawa and H. Nakano, Ferrimagnetism of the Heisenberg models on the quasi-one-dimensional kagome strip lattices, J. Phys. Soc. Jpn. \textbf{81}, 084710 (2012).

\bibitem{ta4}
K. Morita, S. Sota, and T. Tohyama, Resonating dimer–monomer liquid state in a magnetization plateau of a spin-1/2 kagome-strip Heisenberg chain, Commun. Phys. \textbf{4}, 161 (2021).

\bibitem{ta5}
L. Heinze, H. O. Jeschke, I. I. Mazin, A. Metavitsiadis, M. Reehuis, R. Feyerherm, J.-U. Hoffmann, M. Bartkowiak, O. Prokhnenko, A. U. B. Wolter, X. Ding, V. S. Zapf, C. Corvalan Moya, F.Weickert, M. Jaime, K. C. Rule, D. Menzel, R. Valenti, W. Brenig, and S. S$\ddot{\rm{u}}$llow, Magnetization Process of Atacamite: A Case of Weakly Coupled $S$ =1/2 Sawtooth Chains, Phys. Rev. Lett. \textbf{126}, 207201 (2021).

\bibitem{ta6}
M. Sato, Coexistence of vector chiral order and Tomonaga-Luttinger liquid in the frustrated three-leg spin tube in a magnetic field, Phys. Rev. B \textbf{75}, 174407 (2007).

\bibitem{ta7}
X. Plat, S. Capponi, and P. Pujol, Combined analytical and numerical approach to magnetization plateaux in one-dimensional spin tube antiferromagnets, Phys. Rev. B \textbf{85}, 174423 (2012).

\bibitem{ta8}
T. Ito, C. Iino, and N. Shibata, Thermodynamic properties of the $S$ =1/2 twisted triangular spin tube, Phys. Rev. B \textbf{97}, 184409 (2018).

\bibitem{ta9}
T. Sugimoto, M. Mori, T. Tohyama, and S. Maekawa, Magnetic phase diagram of a frustrated spin ladder, Phys. Rev. B \textbf{97}, 144424 (2018). 

\bibitem{ta10}
J. Stre$\check{\rm{c}}$ka, J. Richter, O. Derzhko, T. Verkholyak, and K. Kar$\acute{\rm{l}}$ov$\acute{\rm{a}}$, Diversity of quantum ground states and quantum phase transitions of a spin-1/2 Heisenberg octahedral chain, Phys. Rev. B \textbf{95}, 224415 (2017).


\bibitem{TCNQ_square} 
H. Yamaguchi, Y. Tamekuni, Y. Iwasaki, and Y. Hosokoshi, Candidate for a fully frustrated square lattice in a verdazyl-based salt, Phys. Rev. B {\bf 97}, 201109(R) (2018).

\bibitem{PF6}
H. Yamaguchi, Y. Sasaki, T. Okubo, M. Yoshida, T. Kida, M. Hagiwara, Y. Kono, S. Kittaka, T. Sakakibara, M. Takigawa, Y. Iwasaki, and Y. Hosokoshi, Field-enhanced quantum fluctuation in an $S$ =1/2 frustrated square lattice, Phys. Rev B, \textbf{98}, 094402 (2018).

\bibitem{SbF6}
H. Yamaguchi, Y. Iwasaki, Y. Kono, T. Okubo, S. Miyamoto, Y. Hosokoshi, A. Matsuo, T. Sakakibara, T. Kida, and M. Hagiwara, Quantum critical phenomena in a spin-1/2 frustrated square lattice with spatial anisotropy, Phys. Rev B, \textbf{103}, L220407 (2021).

\bibitem{Zn_gap}
H. Yamaguchi, N. Uemoto, T. Okubo, Y. Kono, S. Kittaka, T. Sakakibara, T. Yajima, S. Shimono, Y. Iwasaki, and Y. Hosokoshi, Gapped ground state in a spin-1/2 frustrated square lattice, Phys. Rev. B \textbf{104}, L060411 (2021).

\bibitem{MnCl4}
H. Yamaguchi, T. Okubo, A. Matsuo, T. Kawakami, Y. Iwasaki, T. Takahashi, Y. Hosokoshi, and K. Kindo, Quantum spin state stabilized by coupling with classical spins, Phys. Rev. B \textbf{109}, L100404 (2024).

\bibitem{zigzag_square}
H. Yamaguchi, K. Shimamura, Y. Yoshida, A. Matsuo, K. Kindo, K. Nakano, S. Morota, Y. Hosokoshi, T. Kida, Y. Iwasaki, S. Shimono, K. Araki, and M. Hagiwara, Field-induced quantum phase in a frustrated zigzag-square lattice, Phys. Rev. Mater. \textbf{7}, L091401 (2023).

\bibitem{a26Cl2V}
H. Yamaguchi, T. Okubo, S. Kittaka, T. Sakakibara, K. Araki, K. Iwase, N. Amaya, T. Ono, and Y. Hosokoshi, Experimental realization of a quantum pentagonal lattice, Sci. Rep. \textbf{5}, 15327 (2015).

\bibitem{Kondo}
H. Yamaguchi, Y. Tominaga, T. Kida, K. Araki, T. Kawakami, Y. Iwasaki, K. Kimura, and M. Hagiwara, Realization of a spin-1/2 Kondo necklace model with magnetic field-induced coupling switch, Phys. Rev. Res. \textbf{7}, L012023 (2025).
 
\bibitem{Doniach}
S. Doniach, The Kondo lattice and weak antiferromagnetism, Physica B \textbf{91}, 231 (1977).


\bibitem{riron1}
T. Yamamoto, M. Asano, and C. Ishii, Magnetization process of one-dimensional Kondo necklace model with next-nearest-neighbor interaction, J. Phys. Soc. Jpn. \textbf{70}, 3678 (2001).

\bibitem{riron2}
T. Yamamoto, K. Ide, and C. Ishii, Phase diagram and critical properties of the frustrated Kondo necklace model in a magnetic field, Phys. Rev. B \textbf{66}, 104408 (2000).

\bibitem{riron4}
W.-L. Tu, E.-G. Moon, K.-W. Lee, W. E. Pickett, and H.-Y. Lee, Field-induced Bose-Einstein condensation and supersolid in the two-dimensional Kondo necklace, Commun. Phys. \textbf{5}, 130 (2022).

\bibitem{verd}
R. Kuhn, $\ddot{\rm{U}}$ber Verdazyle und verwandte Stickstoffradikale, Angew. Chem. \textbf{76}, 691 (1964).

\bibitem{MOcal} 
M. Shoji, K. Koizumi, Y. Kitagawa, T. Kawakami, S. Yamanaka, M. Okumura, and K. Yamaguchi, A general algorithm for calculation of Heisenberg exchange integrals J in multispin systems, Chem. Phys. Lett. {\bf 432}, 343 (2006).

\bibitem{morotaMn}
H. Yamaguchi, S. C. Furuya, S. Morota, S. Shimono, T. Kawakami, Y. Kusanose, Y. Shimura, K. Nakano, and Y. Hosokoshi, Observation of thermodynamics originating from a mixed-spin ferromagnetic chain, Phys. Rev. B \textbf{106}, L100404 (2022).

\bibitem{tominagaNi}
Y. Tominaga, A. Matsuo, K. Kindo, S. Shimono, K. Araki, Y. Iwasaki, Y. Hosokoshi, S. Noguchi, and H. Yamaguchi, Mixed-spin two-dimensional lattice composed of spins 1/2 and 1 in a radical-Ni complex, Phys. Rev. B \textbf{108}, 024424 (2023).

\bibitem{morotaCo}
S. Morota, Y. Iwasaki, M. Hagiwara, Y. Hosokoshi, and H. Yamaguchi, Magnetic anisotropy in a verdazyl-based complex with cobalt(II), J. Phys. Soc. Jpn. \textbf{92}, 054705 (2023). 

\bibitem{Masuda}
T. Masuda, A. Zheludev, B. Grenier, S. Imai, K. Uchinokura, E. Ressouche, and S. Park, Cooperative ordering of gapped and gapless spin networks in Cu$_2$Fe$_2$Ge$_4$O$_{13}$, Phys. Rev. Lett. {\bf 93}, 077202 (2004).

\bibitem{b26Cl2V} 
H. Yamaguchi, T. Okubo, K. Iwase, T. Ono, Y. Kono, S. Kittaka, T. Sakakibara, A. Matsuo, K. Kindo, and Y. Hosokoshi, Various regimes of quantum behavior in an $S$=1/2 Heisenberg antiferromagnetic chain with fourfold periodicity, Phys. Rev. B {\bf 88}, 174410 (2013).

\bibitem{random_ladder} 
Y. Tominaga, I. Shimamura, T. Kida, M. Hagiwara, K. Araki, Y. Hosokoshi, Y. Iwasaki, and H. Yamaguchi, Gapless behavior in a two-leg spin ladder with bond randomness, Phys. Rev. B {\bf 110}, 134430 (2025).

\bibitem{data} 
H. Yamaguchi, Experimental data of ($m$-Py-V)$_3$[Ni(NO$_3$)$_2$] (2025), https://doi.org/...(TBA)











































\end{thebibliography}
\end{document}